\documentstyle[11pt,AATS,epsf]{article}	

\def\ifm#1{\relax\ifmmode#1\else$\mathsurround=0pt #1$\fi}
\def\kms{\ifmmode\,{\rm km}\,{\rm s}^{-1}\else km$\,$s$^{-1}$\fi}
\def\hmpc{\,h\ifm{^{-1}}{\rm Mpc}}
\def\ihmpc{\,(h^{-1} {\rm Mpc})^{-1}}

\def\m3{Mark\ III}

\def\etal{{\it et al.\ }}

\def\eg{{\it e.g.}}

\def\re{\reference}

\def\gtaprx{\mathrel{\vcenter{\offinterlineskip \hbox{$>$}
    \kern 0.3ex \hbox{$\sim$}}}}
\def\ltaprx{\mathrel{\vcenter{\offinterlineskip \hbox{$<$}
    \kern 0.3ex \hbox{$\sim$}}}}
\def\capt{\small \baselineskip 12pt }

\def\pmb#1{\setbox0=\hbox{#1}%
 \kern-.025em\copy0\kern-\wd0
 \kern.05em\copy0\kern-\wd0
 \kern-.025em\raise.0433em\box0}

\def\vV{\pmb{$V$}}

\def\v0{\pmb{$0$}}

\def \nhat{\ifmmode {\hat{\bf n}}\else${\hat {\bf n}}$\fi}
\def \xhat{\ifmmode {\hat{\bf x}}\else${\hat {\bf x}}$\fi}
\def \yhat{\ifmmode {\hat{\bf y}}\else${\hat {\bf y}}$\fi}
\def \zhat{\ifmmode {\hat{\bf z}}\else${\hat {\bf z}}$\fi}

\newcommand{\vev}[1]{\left\langle#1\right\rangle}
\newcommand{\bfv}{{\mathbf v}}

\newcommand{\be}{\begin{equation}}
\newcommand{\ee}{\end{equation}}

\begin{document}	
\title{Cosmic Flows: A Status Report} 

\author{S. Courteau}
\affil{University of British Columbia, Physics \& Astronomy, 
       Vancouver, BC., Canada}
\author{A. Dekel}
\affil{Racah Institute of Physics, The  Hebrew University, 
 Jerusalem 91904, Israel}


\begin{abstract}
We give a brief review of recent developments in the study of the 
large-scale velocity field of galaxies since the international 
workshop on Cosmic Flows held in July 1999 in Victoria, B.C.  
Peculiar velocities (PVs) yield a tight and unique constraint 
on cosmological characteristics, 
independent of $\Lambda$ and biasing, such as the cosmological matter 
density parameter ($\Omega_m$) and the 
convergence of bulk flows on large scales.
Significant progress towards incorporating non-linear dynamics 
and improvements
of velocity field reconstruction techniques have led to a rigorous control 
of errors and much refined cosmic flow analyses.  Current investigations
favor low-amplitude ($\ltaprx 250$ \kms) bulk flows on the largest scales 
($\ltaprx 100 h^{-1}$ Mpc) probed reliably by existing redshift-distance
surveys, consistent with favored $\Lambda$CDM cosmogonies. 
Tidal field 
analyses also suggest that the Shapley Concentration (SC), located
behind the Great Attractor (GA), might play an important dynamical role, even 
at the Local Group.  Low-amplitude density fluctuations on very large scales
generate the overall large-scale streaming motions while massive attractors 
like the GA, and Perseus-Pisces 
account for smaller scale motions which 
are superposed on the large-scale flow. 
Likelihood analyses of galaxy PVs,
in the framework of flat CDM cosmology,
now provide tight constraints of $\Omega_m = 0.35 \pm 0.05$.
A four-fold size increase of our data base is expected in 
$\sim 4-5$ years with the completion of next generation FP/TF surveys and 
automated supernovae searches within 20,000 \kms.  
\end{abstract}

\section{Introduction}
Ever since the discovery of the microwave background dipole
by Smoot, Gorenstein \& Muller (1977) and the pioneering measurements 
of galaxy motions by Rubin \etal (1976), the study of cosmic flows, or 
deviations from a smooth Hubble flow due to large-scale gravitational 
perturbations, has been recognized as one of the most powerful constraints
to cosmological scenarios (Peebles 1980, Dekel 1994, Strauss \& Willick 1995). 
Indeed, under the assumption that cosmic structure originated from 
small-amplitude density fluctuations that were amplified by gravitational 
instability, the peculiar velocity \pmb{v}\ and mass density contrast 
$\delta$ are together linked 
in the linear regime 
by a deceptively simple expression (from
mass conservation in linear perturbation theory): 
\be
  \nabla \cdot \bfv = -\Omega_m^{0.6} \delta.
\ee
The mean square bulk velocity on a scale $R$ is easily calculated in
Fourier space as: 
\be
 \vev{v^2(R)} = \frac{\Omega_m^{1.2}}{2\pi^2}
 \int_0^\infty P(k)\, \widetilde W^2(kR) \, dk \,, 
\ee
where $P(k)$ is the mass fluctuation power spectrum and 
$\widetilde W^2(kR)$ is the Fourier transform of a top-hat window 
of radius $R.$  Measurements of galaxy PVs can thus 
directly constrain $\Omega_m$, the shape and amplitude of the power 
spectrum, and test assumptions about the 
statistical properties of the initial fluctuations and gravitational 
instability as the engine of perturbation growth. 

The last major workshop on Cosmic Flows in July 1999 in Victoria, BC
(Courteau, Strauss, \& Willick 2000; hereafter CFW2000) came at a time 
when important new data sets and critical modeling of the 
``biasing" relation between 
the galaxy and mass distribution
were just being released.  
Fundamental questions debated at the conference, and central to all
cosmological investigations based on cosmic flows, included\footnote{
Discussions about the measurements of the small-scale velocity dispersion 
and the coldness of the velocity field also figured prominently in the
workshop agenda but we do not offer any update below, for lack of space.
The interested reader should read CFW2000.}: (1) {\it What is the 
amplitude of bulk flows on the largest scales probed}? 
(2) {\it Can velocity analysis provide accurate estimates of $\Omega_m$}?, 
and (3) {\it What is the value of $\Omega_m$}? 
The last two years have seen significant progress providing nearly 
definitive answers to each of the 3 questions above, as we discuss 
in the remainder of this review.  

Detailed information about cosmic flows can be found in the {\it Cosmic
Flows 1999} workshop proceedings (CFW2000), including the conference 
review by Dekel (2000). 
Also in Willick (1999) and Dekel (1999), as well as Willick (2000). 

\section{Data Sets and Bulk Flows}

The radial peculiar velocity of a galaxy is derived by subtracting the
Hubble velocity $H_0 d$  from the total velocity (redshift) $cz$
in the desired frame of reference (\eg\ CMB or Local Group).
The distance $d$ is inferred from a 
distance indicator (DI) whose accuracy dictates the range of applicability
of the technique.  The relative distance error of common  
DIs ranges from 20\% (Tully-Fisher [TF], Fundamental Plane [FP], Brightest 
Cluster Galaxy [BCG]) down to 5-8\% (Surface Brightness Fluctuations [SBF], 
SNIa, Kinetic Sunyaev-Zel'dovich [kSZ]). The bulk velocity $\vV_B$ of an 
ensemble of galaxies within a sphere (or a shell) of radius R is computed 
by a least square fit of a bulk velocity model predictions $\vV_B\cdot\nhat$
to the observed radial peculiar velocities, where $\nhat$ is a unit vector 
in the direction of the object.  Current results are summarized in Table 1 
and represented graphically in Figure 1.

\newpage

The data sets can de divided into two groups which lie 
either exactly within or somewhat above the predictions from most 
($\Lambda$)CDM families.  Fig.~1 shows the theoretical prediction 
of a $\Lambda$CDM model for the simplest statistic: the bulk-flow 
amplitude in a top-hat sphere.  The solid line is the rms value,
obtained by Eq.~2. 
The dashed lines 
represent 90\% cosmic scatter in the Maxwellian distribution of $V$, 
when only one random sphere is sampled.  With the exception of BCG,
the directions of the non-zero flow vectors are similar (they all
lie within $30^\circ$ of $(l,b)=(280^\circ,0^\circ)$) and the 
velocity amplitudes can be roughly compared even though the survey 
geometries and inherent sample biases can differ quite appreciably. 
A rigorous comparison of flow analyses must however account for 
different window functions (Kaiser 1988, Watkins \& Feldman 1995, 
Hudson et~al. 2000).  Still, the obvious interpretation of these 
data is that of a gradual decline of the flow amplitude, or 
``convergence'' of the flow field to the rest-frame of the CMB 
at $\sim 100 h^{-1}$ Mpc,  consistent with the theoretical assumption 
of large-scale homogeneity.  

\medskip

Cosmic variance however prevents any convergence to complete rest.  
Some of the reported error bars are based on a careful error analysis
using mock catalogs, while others are crude estimates. In most cases 
they represent random errors only and underestimate the systematic biases.  
Large error bars for surveys such as BCG, LP10, SMAC, SNIa, and Shellflow, 
with fewer than a thousand ``test particles,'' are largely due to sampling 
errors which also increase with increasing volumes. 

\medskip

While present bulk flow estimates are in comforting agreement with current 
cosmologies, important efforts are currently 
underway to reduce the systematic and random errors inherent in most
compilations of galaxy PVs, especially at large 
distance.  The former is addressed by collecting homogeneous data 
across the entire sky, in the spirit of Lauer-Postman and Shellflow 
(Courteau et~al. 2000).  
The latter simply requires that large numbers of galaxies and cluster 
of galaxies be observed to reduce Poisson noise and systematic biases.  
The nominal sample size to achieve a minimum signal/noise for each 
spherical volume chosen must be estimated from mock catalogs based on 
an expected number density profile (as a function of distance or redshift 
from us) and sky coverage. 
New surveys including many thousand ``test'' particles and reaching 
out to 15,000 \kms should quantify the convergence of the peculiar 
velocity field on very large scales.  These surveys include, for 
example, NFP\footnote{\tt astro.uwaterloo.ca/$\sim$mjhudson/nfp/} 
for the FP measurements of $\sim 4000$ early-type galaxies in 
100 X-ray selected clusters,
6dF\footnote{\tt msowww.anu.edu.au/colless/6dF/}
for the FP measurements of $\sim 15,000$ Southern hemisphere early-type 
galaxies, the SNfactory\footnote{{\tt snfactory.lbl.gov}.  The detection 
range should actually extend out to 24,000 \kms.} for the 
serendipitous detection and subsequent follow-up of a few hundred SNe per 
year (Aldering 2001, private communication), and the 
Warpfire\footnote{\tt www.noao.edu/noao/staff/lauer/warpfire/}
extension of Lauer \& Postman (1994)'s BCG analysis.
These studies should be completed by 2005, if not sooner. 
\newpage

\begin{figure*}[ht]	
  \plotfiddle{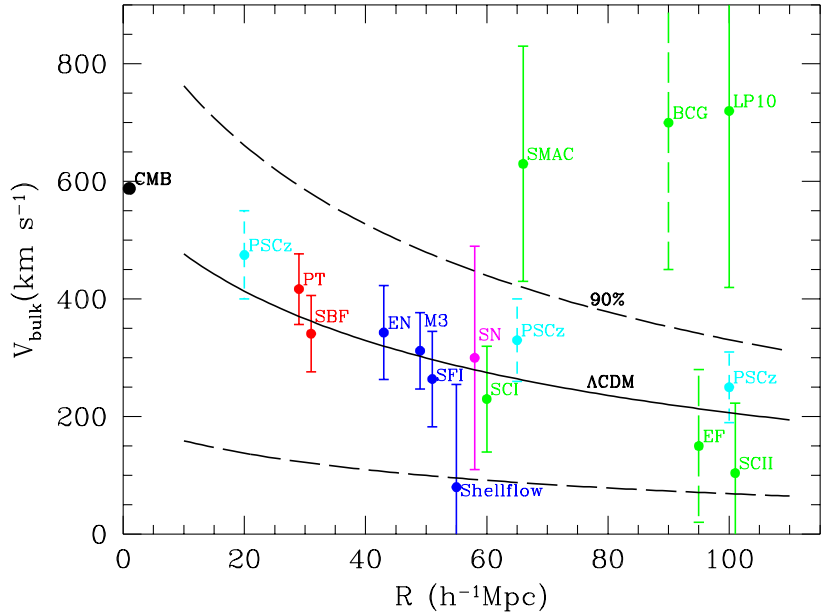}{7.5cm}{0}{60}{60}{-185}{-200}
  \vspace{0.5cm}
 \caption{\protect \capt
Amplitude of CMB bulk velocity in top-hat spheres about the LG,
in comparison with theory.
The curves are the predicted rms and cosmic scatter
for a $\Lambda$CDM model.
The measurements, based on the data listed in Table 1,
are crudely translated to a top-hat bulk velocity.
The error bars are random only.
All the non-zero vectors (except BCG) point to
$(l,b)=(280^\circ,0^\circ)\pm30^\circ$.
Shown as well are the LG dipole velocity (labeled ``CMB''), 
and linear estimates from the PSCz redshift survey for $\beta=0.7$.
Care must be exercised when interpreting such plots since directions 
are not plotted and projected amplitudes ($V_X, V_Y, V_Z$) may 
differ substantially (\eg\ Hudson et~al. 2000).}
\end{figure*}

\begin{center}
\centerline{{\sc Table I. Recent Bulk Flow Measurements}$^\dagger$}
\medskip
\begin{tabular}{l c c | l}\hline\hline
Survey  & $R_{\rm eff}$ ($\kms$) & $V_B$ ($\kms$) & Dist. Ind. \\ \hline
Lauer-Postman \ (BCG)   & 12500 & 700 & BCG \\
Willick \     (LP10K)   & 11000 & 700 & TF  \\
Hudson et~al. \ (SMAC)  &  8000 & 600 & FP  \\ \hline\hline
Tonry et~al. \ (SBF)       & 3000 & 290 & SBF \\
Wegner et~al. \ (ENEAR)    & 5500 & 340 & $D_n$-$\sigma$ \\
Dekel et al. \ (POTENT/M3) & 6000 & 350 & TF,$D_n$-$\sigma$ \\
Riess et~al. \  (SNIa)     & 6000 & 300 & SN Ia \\
Courteau et~al. \ (SHELLFLOW) & 6000 & 70 & TF  \\
Dale \& Giovanelli \ (SFI)    & 6500 & 200 & TF \\
Colless et al. \ (EFAR)       & 10000 & 170 & FP  \\
Dale \& Giovanelli (SCI/SCII) & 14000 & 170 & TF \\ \hline
\end{tabular}
\end{center}
$\dagger$ \ All references in CFW2000.  With the exception of 
Lauer-Postman (1994), all results are post-1999.

\vfill
\newpage

\subsection{The Large-Scale Tidal Field}

The cosmological peculiar velocity field at any point 
can be decomposed into the sum of a ``divergent" field due to density 
fluctuations inside the surveyed volume, and a tidal (shear) field, 
consisting of a bulk velocity and higher moments, due to the 
matter distribution outside the surveyed volume. 
This procedure was carried out by Hoffman et~al. (2001), 
using reconstructions by POTENT (Dekel et~al.~1999) or Wiener Filter
(Zaroubi, Hoffman \& Dekel 1999),
with respect to a sphere of radius $60h^{-1}$ Mpc about the Local group.
Their results are illustrated in Fig.~2.
The divergent component is dominated by the flows into the Great Attractor
(left) and Perseus-Pisces (right), and away from the void in between.
The tidal field shows, for example, that about 50\% of the  
velocity of the Local Group in the CMB frame is due to  
external density fluctuations.  Their analysis suggests the non-negligible
dynamical role of super-structures at distances of $100-200 h^{-1}$ Mpc, 
specifically the Shapley Concentration and two great voids. These should 
be taken into account when considering the convergence of
bulk velocity from different surveys on different scales and of the
dipole motion of the Local Group.

\medskip
\begin{figure*}[b!]	
  \plotfiddle{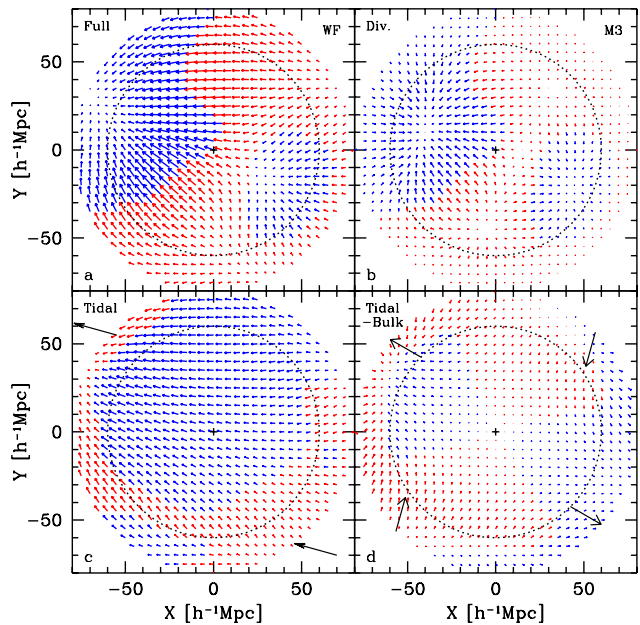}{5cm}{0}{45}{45}{-140}{-150}
  \vspace{3cm}
 \caption{\protect \capt
Wiener filter reconstruction by Hoffman et~al.~(2001)
of the Mark III velocity field in 
the Supergalactic plane, with respect to the sphere of $60\hmpc$ 
about the Local Group (center).
The velocities are measured in $\!\hmpc$ ($1\hmpc=100\kms$).
(a) The full velocity field.
(b) The divergent component due to the mass fluctuations within
    the sphere.
(c) The tidal component due to the mass distribution outside the sphere.
(d) The residual after subtracting the bulk velocity from the tidal component,
    including quadrupole and higher moments.
The black long arrows in the bottom panels show the projected directions
of the bulk velocity and two of the shear eigenvectors respectively.
For more information, refer to Hoffman et~al. (2001). 
}
\end{figure*}

\medskip

\section{Power Spectra and the Measurement of $\Omega_M$}

The peculiar velocities allow direct estimates of $\Omega_m$ independent
of galaxy biasing and $\Lambda$.  Early analyses have consistently yielded 
a lower bound of $\Omega_m > 0.3$ (\eg, Dekel \& Rees 1993), but not 
a tight upper bound.

Cosmological density estimates from 
the confrontation of PVs and the distribution of galaxies in redshift 
surveys have traditionally
yielded values in the range $0.3 < \Omega_m < 1$ (95\% confidence).  This
wide span has often been attributed to nontrivial features of the biasing 
scheme or details of the reconstruction/likelihood method such as the choice
of smoothing length.  Two common approaches to measuring $\Omega_m$ are known
as the {\em density-density\/} (d-d) and {\em velocity-velocity\/} (v-v) 
comparisons.  Density-density comparisons
based on POTENT-like reconstructions (\eg, Sigad et~al. 1998)
have produced typically large values of $\Omega_m$, while v-v 
comparisons yield smaller estimates (\eg, Willick et~al. 1998 [VELMOD], 
Willick 2000, Nusser et~al. 2001, Branchini et~al. 2001).  These differences 
have recently been shown to be insensitive to the complexity of the 
biasing scheme, whether it be non-linear, stochastic, or even non-local 
(Berlind et~al. 2001; see also Feldman et~al. 2001).  Thus, one must look 
for differences inherent to d-d/v-v techniques for an explanation of their
apparent disagreement. 

\medskip

Likelihood analyses of the individual PVs
(\eg\ Zaroubi et~al.~1997, Freudling et~al.~1999, Zehavi \& Dekel 1999)
can be used to estimate the power spectrum of density
fluctuations under the assumption that these are drawn from a Gaussian 
random field.  In linear theory, the shape of the power spectrum $P(k)$ 
does not change with time and thus provides a powerful tool to estimate 
basic cosmological parameters.  Moreover, power spectrum analyses of 
PVs are free of the problems that plague similar determinations from 
redshift surveys such as redshift distortions, triple-valued zones, 
and galaxy biasing, and suffer from weaker non-linear clustering effects. 
Likelihood methods simply require as prior a parametric functional form 
for $P(k)$.  

\medskip

\begin{figure*}[b!]	
  \plotfiddle{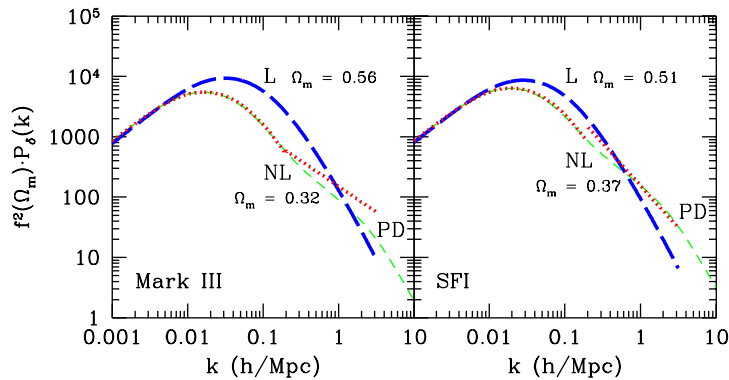}{2cm}{0}{50}{50}{-160}{-150}
  \vspace{2cm}
  \caption{\protect \capt
  The recovered power spectra by the non-linear likelihood analysis of
  Silberman et~al.~(2001) from the data of M3 (left) and SFI (right).  
  The $P(k)$ yielded by the purely linear analysis is 
  marked ``L", while the nonlinear analysis, with a break at $k=0.2\ihmpc$,
  is marked ``NL".  The corresponding values of $\Omega_m$ are marked.
  Also shown for comparison is an extrapolation of the 
  linear part of the recovered $P(k)$ into the nonlinear regime by the 
  Peacock-Dodds (1996) approximation.
  The $P(k)$ is in units of $(\hmpc)^3$.}
\end{figure*}

\medskip

The likelihood analysis of Silberman et~al. (2001) incorporates a correction 
to the power spectrum for non-linear clustering effects, which has been
carefully calibrated using new mock catalogs based on high-resolution
simulations.  The effect of 
this correction, shown in Fig.~3, is to account for larger power on small 
scales and suppress the overall amplitude of $P(k)$
on larger scales where clustering is still linear.  An unbiased fit of
$P(k)$ in the linear regime can thus be achieved, leading to unbiased
constraints on the relevant cosmological parameters. 
The $P(k)$ prior in their analysis assumed a flat $\Lambda$CDM cosmological
model ($h=0.65, n=1,$ COBE normalized), with only $\Omega_m$ as a free
parameter.  Fig.~3 gives final fits based on the
Mark III (Willick et~al. 1997) and SFI (Haynes et~al. 1999) catalogs of
galaxy PVs.  The Mark III catalog is more densely sampled at small 
distances than SFI and also includes elliptical galaxies which are absent 
in SFI; the correction for non-linear effects is thus stronger for Mark III. 
Fitted values for the Mark III data drop from $\Omega_m=0.56 \pm 0.04$ 
in the earlier linear analysis to $0.32 \pm 0.06$ in the improved analysis, 
and for SFI from $0.51 \pm 0.05$ to $0.37 \pm 0.09$.
These revised tight constraints from PVs represent a significant improvement 
in this analysis.

These results are in broad agreement with a recent 
v-v likelihood analysis of SFI PVs against the PSCz IRAS redshift 
survey by Branchini et~al. (2001).

Their procedure entails some
assumptions about the biasing of IRAS galaxies for which PSC redshifts
are measured.  If linear biasing were invoked with a biasing parameter
near unity, Branchini et~al. would find even smaller values of the density 
parameter with $0.15 \leq \Omega_m \leq 0.30$.  This exercise and a direct
comparison with the PV-only likelihood analysis of, say, Silberman et~al.
is 
however futile without a proper prescription of galaxy biasing. 
The direct analysis of PVs by themselves has the advantage of being free 
of the complications introduced by galaxy biasing. 

A $\chi^2$ test applied by Silberman et~al.~to modes of a Principal 
Component Analysis (PCA) 
shows that the nonlinear procedure improves the goodness of fit 
and reduces a spatial gradient that was of concern in the purely linear
analysis.  The PCA allows to address spatial features of the data and to
evaluate and fine-tune the theoretical and error models. 
It demonstrates in particular that the $\Lambda$CDM models used are
appropriate for the cosmological parameter estimation performed.
They also addressed the potential for optimal data compression using PCA,
which is becoming important as the data sets are growing big.

Intriguingly, when Silberman et~al.~allow deviations from $\Lambda$CDM, 
they find an indication for a wiggle in the power spectrum: an excess 
near $k\sim 0.05\ihmpc$ and a deficiency at $k \sim 0.1\ihmpc$ 
--- a ``cold flow".  This may be related to a similar wiggle seen in 
the power spectrum from redshift surveys (Percival et~al. 2001 [2dF]) 
and the second peak in the CMB anisotropy (\eg\ Halverson et~al 2001 [DASI]). 

\section{The Future}

Significant improvements in cosmic flow studies over the 
last couple of years include, for example: 
(1) unbiased recovery of cosmological parameters,
such as $\Omega_m$ and $\sigma_8 \Omega^{0.6}_m$, 
via quasi-nonlinear likelihood analyses of galaxy PVs; 
(2) modeling of non-linear clustering effects in power spectrum analyses 
from PVs, and implementing tools, based on PCA, for evaluating goodness 
of fit; and 
(3) better modeling of biased galaxy formation, in order 
to single out biasing in the comparison of PVs with redshift surveys
and to generate proper mock catalogs for calibrating PV analysis methods.

Future developments rely heavily on growth of the available data 
bases and on refinements of existing catalogs.  The VELMOD technique
has enabled improved recalibrations of the Mark III (Willick et~al. 1998)
and SFI (Branchini et~al. 2001) catalogs
using external information from IRAS redshift surveys. 
We are planning an improved 
recalibration of Mark III using as backbone the homogeneous 
all-sky Shellflow sample, and merging all 
existing catalogs of PVs of field galaxies into a new Mark IV catalog. 

A number of on-going and newly envisioned surveys (6dF, NFP, SNfactory, 
Warpfire) are expected to increase the size of existing data bases by 
a factor 4 within 2005.  New wide-field surveys such as SLOAN, 2MASS, 
and DENIS will also provide most valuable complementary data to help 
control distance calibration errors. 

A noticeable impact to precision flow studies should come from
supernovae searches whose potential to build up very large catalogs of
peculiar velocities (at the rate of a few hundred detections per year) 
and small relative error is unparalleled by no other distance indicator. 
(With ${\Delta d}/{d}({\rm SNIa}) \sim 8\%$, 1 SNIa is worth $\sim 6$ 
TF or FP measurements!)  If a significant fraction of the new SNIe can 
be caught at peak light and monitored to measure a light curve (yielding 
precise distance estimates), current TF/FP data sets will be superseded 
in less than 5 years.  Other ambitious surveys, such as those listed above, 
will complement accurate SN distances with very large data bases thus 
enabling remarkably tight flow solutions in the near future. There are 
good reasons to plan a new workshop on Cosmic Flows in 2005!

\acknowledgments
S.C. would like to thank Ted von Hippel, Chris Simpson, and the scientific 
organizing committee for their invitation, and for putting together a superb 
meeting which was so rich in content and which provided the rare opportunity 
to interact closely with leading (and lively!) scientists from all branches
of astrophysics.  

\medskip

We remain tremendously saddened by the departure of our friend and 
colleague Jeff Willick who did so much for the advancement of cosmic flow
studies and who touched our lives very deeply. 

\newpage


\vspace{1cm}
\hrule
\vspace{0.2cm}

\parindent=0in
\parskip=4mm

Piotr Popowski: A few years ago Avishai Dekel was measuring $\Omega_m=1$
from cosmic flows based on his POTENT method, but now the value you quote
is 0.3. Can you comment on what caused this difference?

St\'{e}phane Courteau: 
Please note that all previous estimates of $\Omega_m$ reported by Dekel
and others based on PVs {\it alone} (via POTENT or other methods) actually
claimed a significant lower bound of $\Omega_m >0.3$ but no tight upper
bounds. 

Their results at the time were consistent with $\Omega_m \sim 1$, but they 
never really claimed a measurement of $\Omega_m=1$. The claimed lower bound 
is still valid, but now with the addition of a significant upper bound, 
ruling out $\Omega_m=1$. The main improvement came from the incorporation of
nonlinear effects in the likelihood analysis, which became possible due
to proper mock catalogs based on high-resolution simulations.

A wider range of estimates has indeed been obtained by comparisons of PV 
data with galaxy redshift surveys.  For example, a ``d-d'' comparison 
by Sigad et~al.~(1998) 
indicated a high value for $\Omega_m$, while ``v-v'' comparisons, such as 
by VELMOD (Willick et~al.~1998), yielded smaller values.  These analyses 
were contaminated by galaxy biasing and nonlinear effects, which gave 
rise to relatively large uncertainties. 


\begin{references}
\re Berlind, A.A, Narayanan, V.K., \& Weinberg, D.H. 2001, ApJ, 549, 688
\re Branchini, E., Freudling, W. Da Costa, L.N., Frenk, C.S.,
           Giovanelli, R., Haynes, M.P., Salzer, J.J., Wegner, G. Zehavi, I. 
           2001, MNRAS, {\it in press} [astro-ph/0104430]
\re Courteau, S., Strauss, M.A., \& Willick, J.A., eds. 2000
           {\em Cosmic Flows 1999: Towards an Understanding of 
	   Large-Scale Structure,} (ASP Conference Series)
\re Courteau, S., Willick, J.A., Strauss, M.A., Schlegel, D., 
           \& Postman, M.\ 2000, ApJ, 544, 636 [Shellflow]
\re Dekel, A.\ 1994, ARA\&A, 32, 271
\re Dekel, A., Eldar, A., Kolatt, T., Yahil, A., Willick, J.A.,  
    Faber, S.M., Courteau, S., \& Burstein, D. 1999, ApJ, 522, 1 [POTENT]
\re Dekel, A.  1999, in Formation of Structure in the Universe,
    eds. A. Dekel \& J.P. Ostriker (Cambridge University Press), 250
\re Dekel, A. 2000, in {\em Cosmic Flows 1999: Towards an Understanding of 
	   Large-Scale Structure}, eds. S. Courteau, S., M.A. Strauss, 
	   \& J.A. Willick (ASP Conference Series), 420
\re Dekel, A., \& Rees, M.J. 1993, ApJL, L1
\re Feldman, H.A., Frieman, J.A., Fry, J.N., Scoccimaro, R. 2001
	   Phys. Rev. Letters, 86, 1434 [astro-ph/0010205]
\re Freudling, W., Zehavi, I., da Costa, L.N., Dekel, A., Eldar, A.,
    Giovanelli, R., Haynes, M.P., Salzer, J.J., Wagner, G., \& Zaroubi, S.
    1999, \apj, 523, 1 \ [Linear likelihood]
\re Haverson, N.W. et~al. 2001 \ [DASI coll.] \ {\it submitted to ApJ} 
          [astroph/0104489]
\re Haynes, M., Giovanelli R., Salzer J., Wegner, G., Freudling W.,
           da Costa L., Herter T., Vogt N. AJ, 1999a, 117, 1668 [SFI]
\re Hoffman, Y., Eldar, A., Zaroubi, S., Dekel, A. 2001, 
           {\it submitted to ApJ} [astro-ph/0102190] 
\re Hudson, M.~J. et~al.~2000, in {\em Cosmic Flows 1999: Towards an 
	    Understanding of Large-Scale Structure}, eds. S. Courteau, S., 
	    M.A. Strauss, \& J.A. Willick (ASP Conference Series),  159 
\re Kaiser, N. 1988, \mnras, 231, 149
\re Nusser, A., da Costa, L.N., Branchini, E., Bernardi, M., Alonso, M.V.,
    Wegner, G., Willmer, C.N.A, Pellegrini, P.S.  2001, \mnras, 320, L21 
    [ENEAR/2dF]
\re Peacock, J.A. \& Dodds, S.J., 1996, MNRAS, 280
\re Percival, W.~J. et~al.~2001 \ [2dF coll.] \ {\it submitted to MNRAS} 
    [astroph/0105252]
\re Peebles, P.J.E. 1980, {\em The Large-Scale Structure of the Universe},
	Princeton University Press 
\re Rubin,~V.C., Thonnard, N., Ford, W.K. \& Roberts,~M.S.
	   1976, AJ, 81, 719. 
\re Sigad, Y., Eldar, A., Dekel, A., Strauss, M.A., \& Yahil, A.  1998, 
    ApJ, 495, 516 \ [d-d]
\re Silberman L., Dekel A., Eldar A., Zehavi I. 2001,  
          ApJ, Aug.~10, {\it in press} [astro-ph/0101361] 
\re Smoot, G.F., Gorenstein, M.V., \& Muller, R.A. 1977, 
	   Phys. Rev. Letters, 39, 898  (see also Corey, B.E.
	   and Wilkinson, D.T, 1976, BAAS, 8, 351)
\re Strauss, M.A., \& Willick, J.A. 1995, Physics Report, 261, 271
\re Watkins, R. \& Feldman, H.A. 1995, ApJ, 453, L73
\re Willick, J.A. 1999, in Formation of Structure in the Universe,
    eds. A. Dekel \& J.P. Ostriker (Cambridge University Press), 213
\re Willick, J.A. 2000, in the (electronic) {\em Proceedings of the XXXVth
           Rencontres de Moriond: Energy Densities in the Universe}
           [astro-ph/0003232]
\re Willick J.A., Courteau S., Faber S., Burstein D., 
           Dekel A. Strauss M. ApJS, 1997, 109, 333  [Mark III]
\re Willick, J.A., Strauss, M., Dekel, A., Kolatt, T. 1998, ApJ, 486, 629 
	   [VELMOD]
\re Zaroubi, S., Zehavi, I., Dekel, A., Hoffman, Y., \& Kolatt, T.  1997, 
    ApJ, 486, 21 [Linear likelihood]
\re Zaroubi, S., Hoffman, Y., \& Dekel, A. 1999, ApJ, 520, 413 [Wiener]
\re Zehavi, I., \& Dekel, A. 1999, Nature, 401, 252
\end{references}
\end{document}